\documentstyle[12pt,epsfig]{article}

\advance\textheight by \topskip
\newcommand{\be}{\begin{equation}}
\newcommand{\ee}{\end{equation}}
\newcommand{\br}{\begin{eqnarray}}
\newcommand{\bea}{\begin{eqnarray}}
\newcommand{\beanon}{\begin{eqnarray*}}
\newcommand{\er}{\end{eqnarray}}
\newcommand{\eea}{\end{eqnarray}}
\newcommand{\eeanon}{\end{eqnarray*}}
\newcommand{\ba}{\begin{array}}
\newcommand{\ea}{\end{array}}
\newcommand{\bi}{\begin{itemize}}
\newcommand{\ei}{\end{itemize}}
\newcommand{\bn}{\begin{enumerate}}
\newcommand{\en}{\end{enumerate}}
\newcommand{\bc}{\begin{center}}
\newcommand{\ec}{\end{center}}

\newcommand{\ar}{\rightarrow}

\newcommand{\Dir}{\kern -6.4pt\Big{/}}
\newcommand{\Dirin}{\kern -10.4pt\Big{/}\kern 4.4pt}
\newcommand{\DDir}{\kern -7.6pt\Big{/}}
\newcommand{\DGir}{\kern -6.0pt\Big{/}}
\newcommand{\qqb}{$ q\bar q$}

\newcommand{\ttb}{$ t\bar t$}
\def\Ord{\buildrel{\scriptscriptstyle <}\over{\scriptscriptstyle\sim}}
\def\sm{\ifmmode{{\cal {SM}}}\else{${\cal {SM}}$}\fi}
\def\mt{\ifmmode{{m_{t}}}\else{${m_{t}}$}\fi}
\def\MH{\ifmmode{{M_{H}}}\else{${M_{H}}$}\fi}
\def\MWpm{\ifmmode{{M_{W^\pm}}}\else{${M_{W^\pm}}$}\fi}
\def\Wpm{\ifmmode{{{W^\pm}}}\else{${{W^\pm}}$}\fi}
\def\pl #1 #2 #3 {{\it Phys.~Lett.} {\bf#1} (#2) #3}
\def\np #1 #2 #3 {{\it Nucl.~Phys.} {\bf#1} (#2) #3}
\def\zp #1 #2 #3 {{\it Z.~Phys.} {\bf#1} (#2) #3}
\def\pr #1 #2 #3 {{\it Phys.~Rev.} {\bf#1} (#2) #3}
\def\prep #1 #2 #3 {{\it Phys.~Rep.} {\bf#1} (#2) #3}
\def\prl #1 #2 #3 {{\it Phys.~Rev.~Lett.} {\bf#1} (#2) #3}
\def\mpl #1 #2 #3 {{\it Mod.~Phys.~Lett.} {\bf#1} (#2) #3}
\def\rmp #1 #2 #3 {{\it Rev. Mod. Phys.} {\bf#1} (#2) #3}
\def\xx #1 #2 #3 {{\bf#1}, (#2) #3}

\begin{document}
\tolerance=100000
\thispagestyle{empty}
\setcounter{page}{0}

\begin{flushright}
{\large RAL-TR-97-066}\\
{\large Cavendish--HEP--97/03}\\
{November 1997\hspace*{.5 truecm}}\\ 
\end{flushright}

\vspace*{\fill}

\begin{center}
{\large \bf 
Top-antitop pair production \\[.25cm]
via $b$-quark initiated processes \\[.25cm]
at the Large Hadron Collider}\\[2.cm]
{\large Stefano 
Moretti\footnote{E-mail: 
moretti@v2.rl.ac.uk}}\\[.3 cm]
{\it Rutherford Appleton Laboratory,}\\
{\it Chilton, Didcot, Oxon OX11 0QX, U.K.}\\[1cm]
\end{center}

\vspace*{\fill}

\begin{abstract}
{\normalsize
\noindent
We study $bt\bar t$ production via subprocesses initiated by
$b$-quarks at the Large Hadron Collider. Both QCD and
electroweak interactions are included in the elementary scattering 
amplitudes for $bg\rightarrow b t\bar t$. 
Since the additional jet 
in the final state (arising from the bottom quark)
is in most cases at very low transverse momentum and very high
pseudorapidity, it tends to escape 
detection. Therefore, such a process can act as a 
background to double-top as well as to single-top channels
exploited in top quark
phenomenology at present and future hadron colliders.
Furthermore, if the additional $b$-jet can be tagged then $bt\bar t$ samples
can be exploited in constraining possible effects of New Physics.
The relevance of this reaction in such contexts  is discussed and
various total and differential rates of phenomenological interest are 
given.}
\end{abstract}

\vspace*{\fill}
\newpage

\section*{1. Introduction} 

The main production channel of top quarks at hadron colliders is that 
proceeding through \qqb\ and $gg$ partonic scatterings \cite{tt}, that is,
via
\be\label{qq}
q\bar q\rightarrow t\bar t,
\ee
and
\be\label{gg}
gg     \rightarrow t\bar t, 
\ee
mediated by QCD, being the electroweak (EW) contributions 
$q\bar q\rightarrow \gamma^*,Z\ar t\bar t$ much smaller\footnote{They
amount to $\approx8\%$ of the total cross section both at the Tevatron and
the Large Hadron Collider (see later on).}.

This mode has been exploited in the recent discovery of the top quark 
at the Tevatron \cite{discovery}. The latest value of the top mass as measured
from the the CDF and D0 experiments is $m_t=175.6\pm 5.5$ GeV \cite{value}, 
which is in very good
agreement with that derived within the Standard Model (SM) from
EW measurements at LEP, SLC,
SPS, Tevatron and in neutrino scattering experiments \cite{theory}:
 $m_t = 177  \pm 7^{+16}_{-19}$ GeV. This clearly represents a great
success of the theory. Furthermore, it is worth recalling that
when the mass of the $W^\pm$ boson will be
determined to a precision of 50 MeV or less (at LEP 2),
the EW theory will also significantly reduce the allowed range for the 
mass of the Higgs boson.

It is clear that the discovery of the top quark and the
measurement of its mass are only the 
first steps. In the near future, various details of the production process 
as well as of the decay channels will have to be thoroughly studied and other
fundamental top parameters (the width, the couplings, etc.) 
will have to be determined.
In this respect, a second mode that will be 
studied at the Tevatron 
is the production channel involving one top
quark only \cite{t}. Although the first mechanism (i.e., 
double-top production) is dominant both at the Tevatron and the Large Hadron 
Collider (LHC) \cite{comparison}, the second (i.e., single-top production)
is very well suited for studying the coupling
between the \Wpm\ boson and the top and bottom quarks \cite{CKM}. 
In fact, the measurement of the
Cabibbo-Kobayashi-Maskawa (CKM) matrix element
$V_{tb}$ cannot be easily performed via \ttb-production and decay.
On the one hand, top-antitop pairs are produced through a `flavour blind'
$gt\bar t$ vertex.
On the other hand, the
dependence on $V_{tb}$ entering in the width of the top is `canceled'
by the fact that branching fraction for the decay $t\ar bW^\pm$ is close
to one in the SM, which predicts 
$0.9989\leq |V_{tb}|\leq 0.9993$ at the 90\% confidence level \cite{CL}.  
In the end, in $t\bar t$ samples,
one can establish a lower limit on $|V_{tb}|$ by studying the 
polar angle of the lepton produced in the top decay channel $t\ar bW^\pm
\ar b\ell\nu_\ell$. In contrast, the single-top production rates are directly
proportional to $|V_{tb}|^2$, so that a simple measurement of total cross
section can lead to a high precision determination of the CKM matrix element
describing the mixing between top and down-type quarks.

The variety of mechanisms which produce
only one top in the final state is very rich. They all involve 
EW interactions. They can be conveniently grouped according to the final state
produced and to the virtual particle content (see Ref.~\cite{boos}), as  
\be\label{tb}
tb-\mbox{production}\qquad\qquad\qquad
pp,p\bar p\ar t \bar b X 
\qquad\qquad~\mbox{(via~$s$-channel~$W^{\pm *}$-bosons)},
\ee
\be\label{tq}
tq-\mbox{production}\qquad\qquad
pp,p\bar p\ar t q X 
\qquad\qquad~\mbox{(via~$t,u$-channel~$W^{\pm *}$-bosons)},
\ee
\be\label{tW}
tW^\pm-\mbox{production}\qquad\qquad
pp,p\bar p\ar t W^\pm X 
\qquad\qquad~\mbox{(via~$s[t,u]$-channel~$b[t]^*$~quarks)},
\ee
where $q$ represents a light (anti)quark, $b$ and $t$ a bottom and a 
top (anti)quark, respectively.

Although the event rate of single-top production at the Tevatron is at present
rather small\footnote{For a review of 
typical single-top cross sections, see, e.g.,
Refs.~\cite{comparison,boos}.}, 
this mode will soon be investigated intensively.
In fact, by 1999 the new Main Injector at the Tevatron should be operating, 
boosting the centre-of-mass (CM) energy $\sqrt s_{p\bar p}$ 
of the colliding proton-antiproton beams from 1.8 TeV of the so-called 
`Run 1' (during the past years 1992--1996) to 2 TeV for the forthcoming `Run 2'
(1999--2001). An increase of the total peak luminosity is also foreseen,
by one order of magnitude, up to $L=2\times10^{32}~{\mbox{cm}}^{-2}~
{\mbox{s}}^{-1}$. A final stage of the Tevatron has also been 
proposed: the `Run 3', always at
$\sqrt s_{p\bar p}=2$ TeV but with the increased luminosity  $L=10^{33}
~{\mbox{cm}}^{-2}~{\mbox{s}}^{-1}$ (thus denominated `TeV33', planned
for the years
2003--2006). Eventually, the $pp$ accelerator LHC at CERN will
take over, starting operations in 2005, with a CM energy of 14 TeV
and a peak luminosity in the range $10^{32}-10^{33}
~{\mbox{cm}}^{-2}~{\mbox{s}}^{-1}$ (Run 1).  

The processes (\ref{qq})--(\ref{tW}) represent the most frequent
production modes of top quarks at the hadronic colliders of the present and 
future generation and they have been extensively studied in the literature.
Many of their higher order corrections have also been calculated 
(mainly from QCD) 
and incorporated in the experimental Monte Carlo programs \cite{corrections}.
It should however be noticed that the accuracy of the 
experimental measurements 
is continuously increasing and the huge amount of data which will be 
collected over the next decade will seriously challenge the current reliability
of the perturbative predictions, so that further theoretical efforts will 
soon be needed in order to match the advances on the experimental 
side. In this respect, it is clear that the increase in luminosity and in
CM energy available will for example mean that rarer production
channels of top quarks could yield
detectable rates at the new generation of hadronic machines.

It is the purpose of this letter to
study the production of top-antitop pairs via $b$-quark initiated processes, 
through the two-to-three body reaction 
\be\label{bg}
bg\rightarrow b t\bar t,
\ee
at both the Tevatron and the LHC.
The Feynman diagrams describing this process are given in
Fig.~1a--b. In our opinion, there are at least three good reasons to expect 
that these events
can eventually become interesting in top physics.
\begin{enumerate}
\item Although the final state of reaction (\ref{bg})
always involves an additional jet with respect to that produced via 
processes (\ref{qq})--(\ref{gg}),
this tends to be at very low transverse momentum and very large 
pseudorapidity. 
Therefore, in most cases, it escapes
detection and the final signature is precisely that of top-antitop production
and decay (\ref{qq})--(\ref{gg}).
\item As the typical (partonic) energy available at supercolliders
increases, the content of $b$-quarks inside the
colliding protons is very much enhanced \cite{EHLQ}, 
this yielding a
much higher probability for $b$-quark initiated scatterings to take place.
\item Contrary to the case of the $t\bar t$ channel,
for which EW contributions are due to only two additional diagrams, for
reaction (\ref{bg}) the number of EW diagrams 
is much larger than that of the QCD ones 
(compare Fig.~1a against Fig.~1b), this 
rendering the EW contributions important in the total cross section.
\end{enumerate}

The impact of even small corrections to $t\bar t$ production and decay
via $q\bar q$-
(dominant at the Tevatron) and $gg$- (dominant at the LHC) scatterings
could be crucial both in studying SM effects
and from the point of view of the searches for 
New Physics. In the first case, $bt\bar t$ events are
interesting as they can represent a background to radiative top decays 
\cite{ttg}, such as $t\bar t(g)\ar b\bar b W^+W^- g$
 (which constitute a testing ground of QCD), as well
as to three-body top decays, such as $t\ar b W^\pm Z$ and especially
$t\ar b W^\pm H$ \cite{bwh} (in which $Z,H\ar b\bar b$).
In the second case, one should recall that, being the mass of the 
top of the order of the EW symmetry breaking scale (the vacuum 
expectation value is in fact 246 GeV), top phenomenology represents
one of the most promising places where to search for phenomena beyond the realm
of
the SM \cite{deviation}. 
Therefore, to assess exactly the amount of all sizable
corrections within the ordinary
dynamics is a benchmark task to be achieved in order
to establish possible deviations from the standard theory.

Among the various possible effects of New Physics which
could be investigated in (both single- and double-) top studies at hadron 
colliders \cite{DSB}--\cite{contact},
we discuss here only two, which are relevant to the case of $bt\bar t$
phenomenology. First, various models for dynamical 
EW symmetry breaking  \cite{DSB}
predict the existence of heavy colour-singlet and 
colour-octet vector states strongly coupled
to the top quark, the so-called `colorons' \cite{coloron}. These 
could produce a resonance decaying into
$t\bar t$-pairs. 
Therefore, the effects of the new particles should be visible in the
spectrum of the $t\bar t$ invariant mass. Second, 
if there are more than six quarks, the CKM matrix element $|V_{tb}|$
could be anywhere between (almost) zero and unity, depending on the
amount of mixing between the third and fourth generation, whereas 
non-universal top couplings differing from the SM
form $V-A$ should appear as a deviation
from the branching ratio of top quarks into $bW^\pm$-pairs with the gauge
bosons polarised longitudinally (which happens around 69\% of the times
in the SM). By studying the angular distributions of the charged lepton
in $t\bar t$-decays \cite{r1214}, one can extract limits on the 
form of the $W^\pm tb$ coupling.

A very last motivation that we put forward to justify our analysis is based
on the observation that the three-vertex process (\ref{bg}) possibly involves
large (compared to the QCD ones, see point 3. above)
contributions with two EW couplings. In some instances these
are both $W^\pm tb$ vertices (graphs 1, 5, 9 and 13 in Fig.~1b), so that 
$|V_{tb}|^4$ terms enter in the total cross section of reaction (\ref{bg}). 
Indeed, we will show that diagram 5 in Fig.~1b is the dominant
EW contribution, as all the other diagrams in Fig.~1b amount to only
$\approx2\%$ of the EW part\footnote{Note that, for reason of gauge invariance,
also diagram 1 must be included together with diagram 5 in the computation,
although this is very much suppressed because of the $g\ar t\bar t$ splitting.
In addition, graphs 9 and 13 are negligible, because of the splitting into 
$tb$-pairs
of a real $W^\pm$.}. Therefore, provided the EW contribution is large enough
in the total cross section (and the QCD one is subtracted), rates 
from process (\ref{bg}) will show a high sensitivity to variations of 
$|V_{tb}|$. 
More in general, all the couplings of the EW model involving a top-quark
appear in the diagrams of Fig.~1b. Therefore, a large variety of
possible deviations from the SM dynamics could be tested in $bt\bar t$ events:
in particular, the existence of a fourth generation of quarks $(u_4,d_4)$ with
a down-like quark $d_4$ with mass around 175 GeV.
In fact, the new particle would reveal itself `directly' in the $\gamma 
d_4\bar d_4$ and $Z d_4\bar d_4$ vertices of the theory.
More important, when a 
$d_4$ flavour is produced instead of a top via process (\ref{bg}), diagrams 
1, 5, 9 and 13 of Fig.~1b do not appear any longer, so that the total cross
section should suffer from a large depletion. In contrast, in single-top
studies via measurements of the $|V_{tb}|$ matrix element the effects of
new flavours could pass unobserved, in case of an extremely small mixing
of the third and fourth generation. 
It will be one of our main concerns to quantify the relevance of the diagrams 
in Fig.~1b with respect to those in Fig.~1a, which only
involve gluon-fermion-antifermion vertices, so that their contribution is
insensitive to the flavour. This can also be
affirmed for pair production of $d_4$-quarks from $q\bar q$ and $gg$-fusion,
the counterpart of processes (\ref{qq})--(\ref{gg}).

In summary, the final aim of 
our study is to assess the relevance of $b t\bar t$ events
as background in both double- and single-top phenomenology 
as well as their own importance in `confirming the identity' of the particle
recently discovered at Fermilab. 
The material we will present has been organised as follows. In the next 
Section we give some details of the calculation
and list the values adopted for the various parameters.
Section 3 is devoted to a discussion of the results.
The conclusions are in Section 4. 

\section*{2. Calculation} 

The tree-level Feynman diagrams that one needs for computing
process (\ref{bg}) are given in Fig.~1a--b.
The pure QCD graphs are displayed in Fig.~1a whereas those involving
also EW vertices are given in Fig.~1b. To calculate the
corresponding amplitudes squared we have used 
MadGraph \cite{tim} and HELAS
\cite{HELAS}. The codes produced have been carefully checked for gauge and BRS
\cite{BRS} invariance. 
The integrations over the appropriate phase spaces have been performed using
{VEGAS} \cite{VEGAS}\footnote{Note that in computing the rates for process
(\ref{bg}) we also have 
included the contribution due to the charged conjugated diagrams of Fig.~1a--b
(i.e., those proceeding via $\bar bg$-fusion).}. 

The $b$-quark in the initial state of reaction (\ref{bg})
has been treated as a constituent of the proton with the appropriate
momentum fraction distribution $f_{b/p}(x,Q^2)$, as given by the 
partonic structure functions. So has been done for the gluon.
However, as the parton distribution functions (PDFs) 
of $b$-quarks inside the proton 
suffer from potentially large (theoretical) 
uncertainties\footnote{In fact, the $b$-sea distributions
are not measured by experiment, rather these are obtained from the gluon
distributions splitting into $b\bar b$ pairs by using the 
Dokshitzer-Gribov-Lipatov-Altarelli-Parisi evolution equations \cite{DGLAP} 
and, in general, the implementation of such dynamics is very
different from set to set in those currently available on the market
\cite{MRSA}--\cite{CTEQ4HQ}.} 
(see, e.g., Ref.~\cite{EHLQ}) 
and those of the gluon 
are not so well known at small $x$,
we have produced our results in the case 
of several different recent next-to-leading structure functions,
such as the sets MRRS(1,2,3) \cite{MRSCHM} and CTEQ4(HQ) \cite{CTEQ4HQ},
which give excellent fits to a wide range of deep inelastic scattering data
(including the measurements from the  HERA $ep$ collider)
and to data on other hard scattering processes. 
In each case the appropriate value of $\Lambda^{(n_f)}_{\overline{{MS}}}$
(in the modified Minimal-Subtraction scheme) has been used.
The QCD strong coupling $\alpha_s$ entering explicitly in the
production cross sections and implicitly in the parton distributions has
been evaluated
at two-loop order, with 
$\Lambda^{(n_f\ne4)}_{\overline{{MS}}}$ calculated according to the 
prescriptions in Ref.~\cite{MARCIANO}
and (in general)
at the scale $\mu=\sqrt{\mathaccent94{s}}$ (i.e., 
the CM energy at partonic level). However, since the choice of the scale
for the structure functions and for $\alpha_s$ represents a source of 
uncertainty, we have studied the yields of process (\ref{bg})
also in case of other representative values of $\mu$.

In the numerical calculations presented in the next Section
we have adopted the following values for the electromagnetic coupling constant
and the weak mixing angle:
$\alpha_{em}= 1/128$ and  $\sin^2\theta_W=0.2320$.
For the gauge boson masses and widths we have taken
$M_{Z}=91.19$ GeV, $\Gamma_{Z}=2.5$ GeV,
$M_{W^\pm}=80.23$ GeV and
$\Gamma_{W^\pm}=2.08$ GeV, while for the fermion masses we have used, in 
general, $m_b=4.3$ GeV (to match the value used in the MRRS(1,2,3) PDFs) and 
$m_t=175$ GeV \cite{value}. We have changed the value of $m_b$ into
5 GeV when using the CTEQ4(HQ) PDFs and adopted the additional 
values 165, 170, 180 and
185 GeV for $m_t$ when studying the effects of a possible fourth generation
of quarks. For simplicity, we have
set the CKM matrix element of the top-bottom coupling equal to one.

The Higgs boson of the SM enters directly in the diagrams of Fig.~1b.
As default value for its mass we have used $M_H=150$ GeV, according to
the best $\chi^2$ fit as obtained from the
analysis of the LEP and SLC high precision EW data: 
i.e., $M_H=149^{+148}_{-82}$ GeV \cite{theory}. However, since the
constraints on the Higgs mass are rather weak (a lower bound of
66 GeV from direct searches \cite{lower} and a 95\% confidence level
upper limit of 550 GeV from the mentioned data  exist \cite{rep}) we have 
 studied
the $M_H$ dependence of the EW contributions of process
(\ref{bg}).

Finally, as total CM energies of the colliding beams at the Tevatron and the 
LHC we have adopted the values $\sqrt s_{p\bar p}=2$ and 
$\sqrt s_{pp}=14$ TeV, respectively.

\section*{3. Results}

As a first result we quote the cross section for events of the type
(\ref{bg}) at $\sqrt s_{p\bar p}=2$ TeV. This is very
very small, around 2--3 fb. Thus, it is negligible both in 
double-top and single-top phenomenology. In fact, at lowest order,
the total cross section of the former is around 8 pb \cite{comparison}
whereas that of the latter is approximately 3 pb \cite{comparison,boos}.
That is, process (\ref{bg}) represents a correction of the order of
one part in ten thousands. 
In fact, rates become even smaller when acceptance cuts (in transverse 
momentum, pseudorapidity and separation of the detectable particles) are 
implemented,
in all possible top-antitop decay channels: hadronic, semi-hadronic(leptonic)
and purely leptonic as well. The number of produced events is in itself
tiny too, a few tens at the most at only during Run 3.
Clearly, this is of no experimental relevance. In contrast, at the
LHC, the total cross section is more than three orders of magnitude larger.
Therefore, in the following we will focus our attention to the case of the
CERN hadron collider only.

Before presenting our results for the LHC a few words are needed concerning
the dynamics of process (\ref{bg}). In particular, we would like to point out
that its cross section is finite over all the
available phase space (when all masses are retained in the calculation).
This allow us to safely calculate the rates of process (\ref{bg})
in both the following cases: (i) when the additional $b$-quark in the final 
state is produced in the detector acceptance region and (ii) when it escapes
detection either because at low transverse momentum or because at large
pseudorapidity. Indeed, from the point of view of top-antitop phenomenology
the latter case is more interesting. In fact, as mentioned in the Introduction,
a reason to study  reaction (\ref{bg}) is that it
represents an irreducible contribution to top-antitop production and
decay when the additional jet goes undetected. The former case is instead 
interesting as it can represent a background to radiative top decays 
and to three-body top decays.

Furthermore, we should also mention that in calculating the differential
distributions involving the top decay products we have in general resorted
 to a
Narrow Width Approximation (NWA), as we have 
written the heavy quark propagator as
\be\label{propagator}
\frac{ p\Dir  + m_t}{p^2-m_t^2+im_t\Gamma}
\left( \frac{\Gamma}{\Gamma_{t}}\right)^{\frac{1}{2}},
\ee
where $\Gamma_{t}$ is the tree-level top width and where the numerical value
adopted for $\Gamma$ has been $10^{-6}$. This way, one is able to correctly 
reproduce the rates for $t\bar t X$ production times the (squared) 
branching ratio $[BR(t\ar bW^\pm)]^2$, as it is done in many of the 
experimental simulations, which indeed do not include finite width effects
\cite{gattoevolpe}. Incidentally, this simplifies considerably the numerical
calculations as one can integrated out the Breit-Wigner 
dependence of the two top quarks, thus reducing by two the
number of variables of the multidimensional integrations.
(We have verified that the spectra of quantities involving the $b$'s,
the leptons and the light jet produced in the top decays suffer little
from NWA effects.) The only exception has been made when plotting the 
distribution in the partonic CM energy, for which the exact results have been
obtained (i.e., $\Gamma\equiv\Gamma_t$).

Finally, we identify the jets in the final state with the partons from
which they originate, we introduce no jet energy smearing and apply all cuts
at parton level.

\subsection*{3.1 Theoretical error}

The rates of process (\ref{bg}) depend strongly on the $b$-quark
parton distribution function inside the proton. On the one hand, such density
is very poorly constrained by the data, as even the extreme case in
which this is neglected the derived parton densities 
(such as  the GRV PDFs \cite{GRV94,GRV})
can adequately fit the available data. On the other hand, there has been 
a lot of theoretical activity in the recent years in the field of heavy quark
distributions (although mainly focused to the case of charm quarks) 
\cite{MRSCHM,CTEQ4HQ,HQ1,HQ2}.
In the latter context, it should be mentioned
that two approaches had dominated the literature in the past years:
the so-called TFN-scheme, where TFN stands for 
{\sl Three-Flavour-Number} (such as
in the GRV sets), and the FFN-scheme, where FFN stands for
 {\sl Four(and Five)-Flavour-Number}
(such as in the MRS and CTEQ
parton densities).
In the first case, only the `light' flavours $\{g,u,d,s\}$ participate
in the parton dynamics inside the proton, and heavy flavours can only be created
in scattering processes (i.e., `flavour creation'). In the 
second case, also the
`heavy' quarks $\{c,b,(t)\}$ are numbered among the initial state partons,
provided the energy scale of the interaction is large enough (though in most
cases the `massive' parton are indeed treated as `massless', once the density
is turned on at the appropriate threshold\footnote{For a comparison 
between  mass-dependent
and mass-independent evolution of PDFs, see Ref.~\cite{EHLQ}.}), so that 
heavy quarks in the final state 
can also be obtained by exciting the corresponding
flavour inside the proton (i.e., `flavour excitation'). 
It turns out that (see, e.g., Refs.~\cite{HQ2}) the TFN-scheme is the most
suitable for the heavy quark component of the PDFs near threshold
whereas well above this regime is the FFN-scheme that should be used.

More recently, a consistent
formulation of heavy flavour dynamics within the perturbative-QCD (pQCD) 
framework has been given, in Refs. \cite{MRSCHM,HQ1,HQ2}. The new treatment
encompasses both the `flavour
creation' and `flavour excitation' mechanisms and is valid
from the heavy quark threshold up to the high energy regime. It reduces to the
two above approaches in the appropriate limits: i.e., to the TFN scenario
when $\mu\sim m_q$ and to the
FFN one if $\mu\gg m_q$ (where $q=c,b$)\footnote{As this new
approach effectively interpolates between the preceeding two,
it is often referred to as the Variable-Flavour-Number (VFN) scheme.}.
Although the three formulations in Refs.~\cite{MRSCHM,HQ1,HQ2} are slightly
different, their approach is basically the same. Furthermore, based on the
improved theory for heavy quark dynamics,
global analyses and new PDF packages have been made available, such
as the MRRS(1,2,3) \cite{MRSCHM} and CTEQ4(HQ) \cite{CTEQ4HQ} sets. 

Preliminary comparisons between the various 
`heavy quark sets' have been performed
in Refs.~\cite{HQ1,HQ2,bar}, though results are not conclusive yet, since
similarly fitted parton distributions in the various schemes are not available
in the literature at the moment, to allow for a consistent study 
\cite{CTEQ4HQ}. Being such a comparison beyond the scope of this paper, we
confine ourselves to the computation of some relevant rates of 
process (\ref{bg}) with the four new sets of parton distributions. 
The total cross section for $bg\ar bt\bar t$, before any acceptance cut, 
is given in Tab.~I for MRRS(1,2,3) and CTEQ4(HQ),
at the LHC. 
Indeed, 
Tab.~I shows that the differences among the four sets of PDFs are
reasonably contained. In fact, if one assumes
as default the MRRS(3) value, the largest difference occurs 
with respect to CTEQ4(HQ) one, about $+10\%$. Rather than an absolute value,
such a difference should be considered as a {\sl lower limit} (for the present
time) on the error affecting the rates of process (\ref{bg}) due to the
PDFs. 

Before proceeding further, it should be noticed that, together with 
process (\ref{bg}), one should also consider another reaction, namely 
\be\label{gg4}
gg\ar b\bar bt\bar t\qquad\qquad\mbox{(via~$g\ar b\bar b$~splitting)},
\ee
which gives the same final
state as the process $bg\ar bt\bar t$ if one 
of the $b$-jets is close enough to the
beam pipe. As a matter of fact, the two scattering processes (\ref{bg})
and (\ref{gg4}) (in which one of the two incoming 
gluons splits into $b\bar b$ pairs) should be 
considered simultaneously and their cross sections summed up with a subtraction
of a common part in order to avoid double counting
(see, e.g., Ref.~\cite{CTEQ4HQ,HQ1}). That is, from the sum of 
the `flavour excitation' scattering (\ref{bg}) and the gluon-fusion mechanism
(\ref{gg4}), one has to subtract the piece due
to the configurations in which the quark in the 
internal $b$-line (from the $g\ar b\bar b$ splitting) in (\ref{gg4}) 
is on-shell and
collinear to the incoming gluon.
One would
expect that the total results for the combination of $bg\ar bt\bar t$ and 
$gg\ar b\bar bt\bar t$ events should be less $\mu$-scale dependent then the 
rates for process (\ref{bg}) alone \cite{HQ1}. As a criteria
to decide whether the additional $b$-jet falls inside the detectable
region (thus mimicking a two-to-three body hard scattering) or not,
we use the cut $p_T(b)\le 20$ GeV on either of the bottom quarks
(see eq.~(\ref{cuts}) below). 

The total rates due to processes (\ref{bg}) and (\ref{gg4}) with the
mentioned subtraction as a function of the 
$\mu$-scale are reported in the left-hand column of Table.~II.
From there, one can argue that the error associated with the scale
dependence of reaction (\ref{bg}) should be below 30\% or so.
In fact, this is the difference between the value of the cross section, say,
at $\mu=2m_t$ and at 1 TeV. Such numbers
correspond to the case of the MRRS(3) PDFs (our default in the following), 
though we have verified that
similar rates also occur for the other three sets adopted here.

Finally, for completeness, we also have  calculated the contribution
from events of the type 
\be\label{gg5}
gg\ar b\bar bt\bar t\qquad\qquad\mbox{(via~$g\ar t\bar t$~and~$g\ar 
gg$~splittings)},
\ee
in which one of the gluons splits into either $t\bar t$ or $gg$ pairs,
with one of the final state $b$'s missed along the beam line. As a matter
of fact, also such events contribute to the total rate for $bt\bar t$ 
production, on the same footing as the (\ref{gg4}) ones do. The corresponding
cross sections (with one of the $b$'s having 
$p_T(b)\le 20$ GeV) as a function of the $\mu$-scale can be found in 
the right-hand
column of Tab.~II (i.e., in brackets). They amount 
to roughly 18\% of the total rates in the nearby column.

For sake of simplicity in the numerical calculations, 
in the remainder of the paper we will  
only consider the rates of the process $bg\ar bt\bar t$, thus ignoring
for the time being
its interplay with those of reactions (\ref{gg4}) and (\ref{gg5}).

\subsection*{3.2 $bg\ar bt\bar t$ phenomenology}

Contrary to the case of the Tevatron, at the LHC
effects due to process (\ref{bg}) can be
perceptible in various instances. For starting, the cross section is rather 
large per se, as it amounts to 8 pb. Therefore, with 10 fb$^{-1}$ of Run 1
at the LHC, it yields some 80,000 $bt\bar t$ events per year. In one selects 
semi-(hadronic)leptonic SM decays, then the event rate is around 24,000.

It is interesting to separate the total cross section given previously
into its pure QCD component (i.e., diagrams in Fig.~1a) and that
involving EW interactions as well (i.e., diagrams in Fig.~1b). This
is done in Fig.~2, for the LHC.
In addition, to establish whether the presence of the Higgs graphs 4, 8, 12 and
16 (in Fig.~1b)
has any influence on the total cross section we vary $M_H$ in the range
between 60 and 500 GeV. As one might expect such contributions to be 
particularly relevant near the $H\ar t\bar t$ threshold (see diagrams 4 and 16
in Fig.~1b), we have enlarged the interval 340 GeV $\Ord M_H\Ord$ 360 GeV
in the insert in the middle of the plot. As a matter of fact, the 
effect is visible in the EW contribution but imperceptible in the total
cross section.
Thus, Higgs diagrams are of no special concern, whichever the actual value
of $M_H$ is. This implies that no further theoretical uncertainty related
to the unknown parameter $M_H$ enters in our results.

The important feature in Fig.~2 is that the EW contribution is rather large
in the total results. At the LHC and for $M_H=150$ GeV this is around 3 pb,
yielding some 30,000 events during Run 1 (which become 9,000 after implementing
the semi-hadronic(leptonic) SM branching ratios of the two top quarks).
This is around 40\% of the total rate of process (\ref{bg}).
If a fourth generation down-type flavour $d_4$ of mass $m_{d_4}\approx 175$
GeV is instead produced, these rates change dramatically. As previously 
mentioned, this is mainly due to the absence of the dominant EW
diagrams 1 and 5 
in Fig.~1b. However, also 
the different EW couplings between neutral gauge bosons
(i.e., $\gamma$ and $Z$) are responsible for differences. Tab.~III
shows the total cross section (pure QCD graphs plus the EW ones) for 
events of the type $bg\ar bd_4\bar{d}_4$ against that of
$bt\bar t$, for five values of $m_{d_4}$
(conservatively compatible with the latest measurements of $m_t$).
One can see that the rates are very different. Indeed, those
for $bd_4\bar d_4$ production practically coincide with the QCD contribution
of process (\ref{bg})
only (compare the rates in Tab.~III with those in Fig.~2, when $m_t=175$ GeV).
This pattern remains unchanged for all masses considered.

Clearly, if one can isolate top-antitop events compatible
with an additional $b$-quark
in the final state, then it 
could be possible to
search for  deviations such as those revealed in Tab.~III.
This could be achieved in the semi-hadronic(leptonic) top decay channel
by requiring two $b$-tags and one high $p_T$ lepton and by reconstructing
the resonant fermion mass via a three-jet system. The requirement of
two vertex tags imposes an additional reduction of the detected rates. 
By assuming an efficiency of $60\%$ per fiducial $b$-jet \cite{CMS,ATLAS},
the overall one to tag two of these would then be $36\%$. Of the
original $bt\bar t$ events, some 8,000+3,000 would survive 
during Run 1, further reduced by a factor of 9 
after the following detector 
acceptances (on two $b$-jets)\footnote{In parentheses is the 
$p_T$ threshold for charged leptons used as trigger.}:
$$
|\eta(b)|<2, \qquad\qquad\qquad p_T(b)>20~\mbox{GeV},
$$
$$
|\eta(\ell)|<2.5, \qquad\qquad\qquad p_T(\ell)>10(20)~\mbox{GeV},
$$
$$
|\eta(j)|<4, \qquad\qquad\qquad p_T(j)>20~\mbox{GeV},
$$
\be
\label{cuts} 
{p\Dir}_T>20~\mbox{GeV}.
\ee
The final numbers are approximately 880 and 330 for the final
rates from the diagrams in Fig.~1a--b, respectively. This means
that roughly 1200 events are expect within the SM from $bt\bar t$ production
in the detection channel $bb\ell jjj$, where $\ell=e,\mu$ and $j$ is a
jet. Because of the dominance of diagrams involving two $W^\pm tb$ 
couplings in the EW component of $bt\bar t$ events, several
hundred of $bt\bar t$ events will show  a $|V_{tb}|^4$ dependence, which
could possibly be exploited in experimental analyses to furnish an
independent new measurement of the CKM matrix element.  
A fourth generation $d$-type quark $d_4$ with mass similar to that
of the top would only produce the equivalent of the QCD component of
$bt\bar t$, that is, 38\% less events. Finally, we have verified 
that no intrinsic difference exists in the kinematics of the QCD and EW
components in the spectra of the lepton and jets which could be profitably
exploited to separate  these two parts of the cross section.

\subsection*{3.3 Double- and single-top phenomenology}

The numbers in Tab.~I and Fig.~2 certainly compare
rather poorly to the total rates of both double- and single-top events. 
In fact, the cross section for process (\ref{bg})
amounts to $1\%$ of the former and $2\%$ of the latter. 

In the first case,
this clearly implies that effects due to events (\ref{bg}) can hardly be 
disentangled in $t\bar t$ phenomenology. Indeed, one should recall that
the total cross section for pair production of top quarks is made up
by the two components (\ref{qq})--(\ref{gg}) only, each of these yielding
the same signature and each of these with rates much larger than those 
of (\ref{bg}). Of the 800 pb at the LHC, 730 pb come from $gg$- and 70 
from $q\bar q$-fusion \cite{comparison}.  
However, if particular kinematic configurations are selected to reduce
the leading order $t\bar t$ rates (e.g., to disentangle some New Physics
or when studying the mentioned radiative and/or three-body top decays), 
then $bt\bar t$ events could become important.

Fig.~3 (left-hand side)
plots the invariant mass distributions
of the system $t\bar tX$ for events of the type (\ref{qq})--(\ref{gg})
and (\ref{bg}). From there it is clear that if
one looks for inclusive signals (i.e., $X$ represents
any additional particle in the final state) and studies the total invariant
mass of the final state, 
then a `resonance' (dashed curve) should appear
some 180 GeV above the $t\bar t$ threshold at $2m_t$. Furthermore,
 note that also the peak in the
$M_{t\bar t}$ spectrum produced by events of the type (\ref{bg})
(dotted line) is shifted by several GeV (upwards) with respect to that 
due to $q\bar q$- and $gg$-fusion (solid line).
  
The cosine of the polar angle of the (anti)lepton from the
two top decays (signal of a possible contamination
of $V+A$ coupling in the $W^\pm tb$ vertex) in $bt\bar t$ samples 
is significantly different from that originated by processes 
(\ref{qq})--(\ref{gg}),
as can be appreciated in right-hand side of Fig.~3. As the coefficient
of a possible $V+A$ chirality violating current could well be at the
level of few percent compared to that of the  $V-A$ term, it is important
to recognise and subtract the contribution of $bt\bar t$ events from the
experimental sample.

If one compares the rates due to process (\ref{bg}) with the yield
of each of the subprocesses contributing to single-top production,
one discovers that the 400 pb cross section (including all
charged conjugated processes) for $pp\ar tX$ ($X$ does not include 
here a second $t$-quark) 
is built up in the following way \cite{comparison}: $pp\ar tqX$ yields
253 pb, $pp\ar tW^\pm X$ produces 128 pb whereas $pp\ar tbX$ gives 19 pb
\cite{comparison}. 
Therefore, the rates of process (\ref{bg}) are similar
to those for single-top production in association with a bottom quark.

The latter process has a particular relevance in single-top phenomenology.
In fact, there are three parton subprocesses yielding final states of the type,
say, $t \bar bX$ (i.e., for top-antibottom pairs). 
Namely, (i) $q'\bar q\ar t\bar b$; (ii) 
$q'g\ar t\bar b q$; (iii) $q'\bar q  \ar t\bar b g$ (see 
Ref.~\cite{boos})\footnote{Note that in case (ii) only 
graphs with gluons coupled to light quarks are considered, as 
those involving $g\ar b\bar b$ and $g\ar t\bar t$ splitting form a separate
gauge invariant set and need to be considered apart, because of the definition
of the $b$-sea quarks in the partonic distributions (see discussion in 
Ref.~\cite{boos}).}.
Among these, the `ideal' one to be exploited in phenomenological
studies would be $q'\bar q\ar t\bar b$ \cite{SSD}. This is because
the corresponding
cross section can be reliably calculated, for several reasons.
First, the light (anti)quark
distribution functions inside the proton are evaluated at moderate values
of $x$ where they are well known. Second, the QCD corrections
to this process can be calculated easily (they plug into a pure EW process)
up to the order ${\cal O}(\alpha_s^2)$. Those from the initial state
are taken into account by simply constraining the quark-antiquark flux
via a measurement of the rates for $q'\bar q\ar \ell\nu_\ell$. Those from the
final state can be incorporated without ambiguities as there are
no collinear and soft singularities. As a matter of fact, the two radiations
do not 
interfere until the ${\cal O}(\alpha_s^2)$ order.
Furthermore, the kinematics is that of a two-to-two body process, whereas
cases (ii) and (iii) involve an additional light jet in the final 
state, thus case (i) is easier to reconstruct experimentally.
More in general, the process  $q'\bar q\ar t\bar b$
can boast the same advantages with
respect to any of the other single-top production channels.
For one or more of the following reasons. Firstly, the latter
always involve either a gluon, a $b$ or both in the initial 
state\footnote{The only exception is $q\bar q\ar t W^- \bar b$, which is
however two orders of magnitude smaller \cite{boos} and with a more 
complicated final state.}. Secondly, they proceed via QCD interactions.
Thirdly, singularities can occur at higher orders in $\alpha_s$. Therefore, 
in the first case,
they suffer from larger uncertainties due the rather unknown corresponding
PDFs and, in the second and third cases, the inclusion of higher order QCD 
effects
is in general much less trivial\footnote{This is also true for the simple
two-to-two EW process $q'b\ar qt$, where collinear and infrared divergences 
along the massless $q'q$ line spoil the advantages from the factorisation
of the strong corrections at the order ${\cal O}(\alpha_s)$.}.
 
The cross section for $q'\bar q\ar t\bar b$ at the LHC, before any 
acceptance cuts, is around 9 pb \cite{SSD} (including C.C. processes),
almost the
same as that of process (\ref{bg}), thus the latter must be 
considered as possible dangerous source of background events. To establish
its relevance in this context, we proceed as follows. First, we calculate
the cross section of events of the type (\ref{bg}) in which the additional
$b$-quark in the final state escapes detection. In this case, the final
state recorded in the detectors is $t\bar t$. Then, we select
semi-leptonic(hadronic) decays $t\bar t\ar
b\bar b W^+W^-\ar b\bar b \ell\nu_\ell jj$, where $\ell=e,\mu$ and $j$
represents a light quark jet, originated by the fragmentation of
a $u$-, $d$-, $s$- or $c$-quark. To appreciate the kinematics
of the additional final state $b$-quark in reaction (\ref{bg}) one can refer
to Fig.~5, where its distributions in
transverse momentum $p_T$ and pseudorapidity $\eta$ are plotted,
along with those of all the other final state particles, as they would
appear at the LHC\footnote{Note that we also have  
plotted (in Fig.~4) the 
same rates for the Tevatron, although we have already assessed that they are
of no relevance at such a collider. We have done so simply to
show quantitatively that the percentage of events (\ref{bg}) with 
the first $b$-quark produced escaping detection because of the finite coverage
in $p_T$ and $\eta$ is much higher at larger energies: compare the solid
curves in the two plots on the left hand side of Figs.~4 and 5. In contrast, 
the detector requirements on the other particles in the final state are
equally effective at both Tevatron and LHC energies (dashed and dotted
lines in the left windows and all curves in the right ones).}. 
We introduce here a distinction between the three $b$'s in the final
state of (\ref{bg}), that we will maintain in the following. This because
they have different kinematic behaviours. We call `prompt $b$' that 
produced in association with the $t\bar t$ pair, `direct $b$' that 
coming from the decay of the top in Fig.~1a--b 
and `indirect $b$' the remaining one from the antitop\footnote{The
labelling works the other way around for the last two $b$'s if
$\bar bg$ induced diagrams are considered. Note that in the caption of 
Figs.~4--5 we have applied the distinction between `direct' and `indirect'
also top the (anti)top decay products.}.
Such a distinction has phenomenological relevance for the `prompt $b$' for
the key r\^ole that it plays in the determination of the observed rates
of process (\ref{bg}), whereas in the other two cases it
has been made for sake of illustration only. In the first case,
one can appreciate the striking behaviour of the parton in Fig.~5, as this
tends to be emitted almost exclusively along the beam pipe, with $p_T<20$ GeV 
and $|\eta|>4$. For the other two $b$'s, one should recognise that, although 
the production dynamics of the `direct $b$' and the `indirect $b$' is
different (at least in the diagrams of Fig.~1b, not in those
of Fig.~1a) and this can be spotted in the left hand side plots of
Fig.~5 (and 4 too), the differences cannot possibly be 
disentangled experimentally. 
We shall now proceed by performing the same selection procedure
outlined in Ref.~\cite{SSD} and will compare the yield of process
(\ref{bg}) with those presented there.

The acceptance cuts implemented to simulate the detectors are those
already mentioned in eq.~(\ref{cuts}), supplemented by the further requirements
of separation (in general, on the `direct' and `indirect' $b$'s):
\be
\label{cuts1} 
|\Delta R_{bb}|>0.7, \qquad\qquad\qquad |\Delta R_{b\ell}|>0.7. 
\ee
Note that in applying cuts (\ref{cuts})--(\ref{cuts1}) to process
(\ref{bg}) in the context of single-top phenomenology one only has
four jets in the selected final state (two $b$-jets and
two light flavour ones), as one of the three produced $b$-quarks 
(in general, the `prompt' $b$) is assumed
to be lost along the beam pipe.
The cross section of $bg\ar bt\bar t$ events after the above 
acceptance cuts is approximately 1600 fb. After implementing the branching
ratio 2/9 for electronic and/or muonic decays of the tagged $W^\pm$ one
gets $\sigma\approx360$ fb, thus 3,600 events during Run 1 at the LHC
(with ten inverse femtobarns).
The rate obtained for process $q'\bar q\ar t\bar b$
in Ref.~\cite{SSD}  after the same requirements and in the same detection
channel is (reading from Tab.~1 in that paper) $580$ fb (including
charge conjugation).
The signal is thus a factor of approximately 1.6 above the background from 
process (\ref{bg}). The application of the cut $M_{b\bar b}>110$ GeV (see again
Ref.~\cite{SSD}) imposes a factor of two of reduction on $bt\bar t$ events.
Finally, we obtain an additional drastic rejection 
against the latter, because of the additional $W^\pm$ in the final state.
In fact, as can 
be appreciated in Fig.~5, any additional light flavour jet 
$j$ in reaction (\ref{bg}) falls inside the acceptance region defined
by $p_T(j)>20$ GeV and $|\eta(j)|<4$ \cite{SSD}. We estimate that 
the final cross section for $bt\bar t$ events,
after light flavour jet rejection
and after including vertex tagging efficiency would be of a few
femtobarns, that is one order of magnitude smaller than
the signal and two orders smaller than
the background rates obtained from $W^\pm b\bar b$, $W^\pm jj$,
$W^\pm Z$, $t\bar b j$ and $t\bar t$ background events \cite{SSD}.
Therefore, also in single-top SM phenomenology
events of the type (\ref{bg}) are well under control, though they represent
a sizable correction (of several percent) to the $q'\bar q\ar t\bar b$ signal,
which should be considered when searching for New Physics.

\section*{4. Summary and conclusions}

In this paper we have studied top-antitop production in association
with an additional $b$-quark at the Tevatron and the LHC.
The interest in this process for top studies at hadronic colliders
comes from the fact that the $b$-quark is in most cases at low transverse
momentum and large rapidity, so it tends to escape the acceptance region
of the detectors. Therefore, such events enter naturally in the candidate
$t\bar t$ sample and should then be considered if one wants to carry out
detailed studies of top quark properties. 

We have shown that at the Tevatron the total cross section for such events
is negligibly small, around 2--3 
fb for the upgraded Fermilab $p\bar p$ collider
with $\sqrt s_{p\bar p}=2$ TeV. In contrast, at the CERN accelerator
the corresponding number is 8 pb.
This should yield about 80,000 events after Run 1
for an accumulated luminosity of 10 inverse femtobarns. Though such a number 
is certainly affected by a large indetermination due to the $b$-structure
functions, we believed it 
reasonably large anyway so  to consider in detail the relevance
of $bt\bar t$ events at the LHC, both
as a background to double- and single-top events and as a signal on its own.
As for the error associated to the parton density of $b$-quarks, we have
given an estimate of its {\sl lower limit}, of about $10\%$, 
by comparing the total rates as obtained from four very recent sets of 
parton distributions, all implementing the heavy quark dynamics in the 
context of the newly developed Variable-Flavour-Number 
(factorisation) scheme. Furthermore, we have established a $30\%$ uncertainty
of the total rates, depending on the choice of the factorisation scale $\mu$,
which was varied between twice the top mass and the TeV scale.
(Indeed, to minimise the impact of the two mentioned errors, we
have neglected the absolute normalisation of the relevant differential rates.)

In the case of double-top physics, it has been shown that the total rate of 
$bt\bar t$ events is small, as it represents a correction of 1\% only to the 
QCD signal via $q\bar q,gg\ar t\bar t$. However, this effect could be above
the experimental accuracy and further enhanced by dedicated selection 
procedures
(especially in order to disentangle possible effects due to New Physics),
so that it should be considered when proceeding to 
MC simulations.  In particular, 
the kinematics of the top quarks in $bt\bar t$ events is very
different from that of the leading $t\bar t$ production: for example,
 in the $M_{t\bar tX}$ invariant
mass, where $X$ represents any additional particle in the final state.
In fact, the three-jet mass 
distribution $M_{t\bar t b}$ is much broader then the $M_{t\bar t}$ one
and the peak of the resonance is shifted by +180 GeV whereas the two-jet
one $M_{t\bar t}$ is indeed 
similar to that of $t\bar t$ events but the shape of the
threshold resonance is distorted.
Finally, the spectra in the polar angle of the 
decay leptons differ significantly from those produced by ordinary $t\bar t$
events. 

In the case of single-top physics, it has been demonstrated that the
$bt\bar t$ production
rates are similar  to the yield of the process 
$q'\bar q\ar t\bar b$, which has been advocated as the best channel
to probe the $W^\pm tb$ vertex of the underlying theory. Once an appropriate
selection strategy of $t\bar b$ events is adopted, the $bt\bar t$ ones are
however greatly reduced. Nonetheless, the latter 
 should anyway be  included in the experimental simulations
aiming to measure the Cabibbo-Kobayashi-Maskawa matrix element $|V_{tb}|$
and to test the vector/axial structure of the top-bottom-$W^\pm$ coupling, 
as violations of the SM dynamics could well be at the same level and
show the same kinematic features
as those due to  the $bt\bar t$ corrections. 

As for signal on its own, events of the type $bg\ar bt\bar t$ have been proved
to be extremely sensitive to the possible presence of a fourth generation of
quarks, involving a down-type fermion $d_4$ with mass similar to that of the 
top, which could then mimic the latter. This is due to the fact that the EW 
contribution (dependent on the flavour of the produced fermion) is almost 40\% 
of the total $bt\bar t$ rates
and almost coincides
 with diagrams involving off-shell
 $W^\pm$-currents, which are
naturally absent in case of $b d_4\bar d_4$ production. Thus, if the 
measured cross
section suffered from a large depletion, this would indicate that
the heavy particle detected recently at the Tevatron is indeed a new flavour
of the theory. Conversely, about 40\% of the total $bt\bar t$
Standard Model cross section 
shows a $|V_{tb}|^4$ dependence, so that top-antitop events accompanied
by an additional $b$-quark could be exploited to obtain a new independent 
measurement of this crucial quantity.

\section*{5. Acknowledgements}

We are grateful to the UK PPARC for financial support.

\goodbreak

\vfill
\newpage

\subsection*{Table Captions}
\begin{description}

\item{[I]    } Total cross sections (QCD and EW summed) for process
(\ref{bg}) at the LHC  for four different
sets of structure functions. 
Errors are as given by {VEGAS}.

\item{[II]  } Total cross sections (QCD and EW summed) for process
(\ref{bg}) at the LHC  for four different choices of
the scale parameter. In parentheses are the corresponding rates
for process (\ref{gg5}).
The MRRS(3) structure functions have been used.
Errors are as given by {VEGAS}.

\item{[III] }Total cross sections (QCD and EW summed) for processes
$bg\ar bd_4\bar d_4$ (see the text) and (\ref{bg}) at the LHC 
for five different masses $m_t/m_{d_4}$.
The MRRS(3) structure functions have been used.
Errors are as given by {VEGAS}.

\end{description}

\vfill
\newpage

\subsection*{Figure Captions}
\begin{description}

\item{[1]   } Lowest order Feynman diagrams describing process
(\ref{bg}): ({\bf a}) the ${\cal O}(\alpha_{s}^3)$ contribution;
            ({\bf b}) the ${\cal O}(\alpha_{s}\alpha_{em}^2)$ contribution.
The package MadGraph \cite{tim} has been used to produce the PostScript file.
In ({\bf b}) `A' represents a photon and the dashed line identifies the
SM Higgs boson.

\item{[2]   }  Cross sections for process (\ref{bg})
at the LHC as a function of the Higgs mass.
The MRRS(3) structure functions have been used.
Solid line: process (\ref{bg}),
            ${\cal O}(\alpha_{s}^3)+{\cal O}(\alpha_{s}\alpha_{em}^2)$
            rates.
Dashed line: process (\ref{bg}),
             ${\cal O}(\alpha_{s}\alpha_{em}^2)$
             rates.

\item{[3]   } Left window: 
differential distributions in invariant mass of the systems 
$ t\bar t$ in events of the type (\ref{qq})--(\ref{gg}) (solid line),
$bt\bar t$ in events of the type (\ref{bg})             (dashed line),
$ t\bar t$ in events of the type (\ref{bg})             (dotted line).
              Right window: 
differential distributions in polar angle of the 
lepton in events of the type (\ref{qq})--(\ref{gg}) (solid line),
antilepton in events of the type (\ref{qq})--(\ref{gg}) (dashed line),
lepton in events of the type (\ref{bg})                 (dotted line),
antilepton in events of the type (\ref{bg})             (dash-dotted line).
Normalisations are to unity.

\item{[4]   } Differential distributions in transverse momentum
(upper two plots) and in pseudorapidity (lower two plots) of the
particles in the final state of process (\ref{bg}) at the Tevatron.
All QCD and EW
contributions have been considered. On the left hand side, spectra
of the three $b$-quarks: `prompt $b$' (solid line); `direct $b$'
(dashed line) and `indirect $b$' (dotted line).
On the right hand side, spectra of the leptons and light quark
jets: `direct leptons/jets' (solid and dotted lines) and
`indirect leptons/jets' (dashed and dot-dashed lines).
Each curve is normalised to unity.
The MRRS(3) structure functions have been used.

\item{[5]   } Same as Fig.~4 at the LHC.

\end{description}

\vfill
\newpage

\begin{table}
\begin{center}
\begin{tabular}{|c|c|}
\hline
\multicolumn{2}{|c|}
{\rule[0cm]{0cm}{0cm}
$\sigma_{tot}$ (fb)}
\\ \hline
\rule[0cm]{0cm}{0cm}
PDFs & $\sqrt s_{pp}=14$ TeV \\ \hline\hline
\rule[0cm]{0cm}{0cm}
MRRS(1)  & $8117\pm25$ \\
MRRS(2)  & $8135\pm26$ \\
MRRS(3)  & $8101\pm24$ \\
CTEQ(4HQ) & $8996\pm23$ \\ \hline\hline
\multicolumn{2}{|c|}
{\rule[0cm]{0cm}{0cm}
$M_H=150$ GeV}
\\ \hline

\multicolumn{2}{|c|}
{\rule[0cm]{0cm}{0cm}
no acceptance cuts}
\\ \hline

\multicolumn{2}{c}
{\rule{0cm}{1.0cm}
{\Large Tab. I}}  \\

\end{tabular}
\end{center}
\end{table}

\vfill
\newpage 

\begin{table}
\begin{center}
\begin{tabular}{|c|c|}
\hline
\multicolumn{2}{|c|}
{\rule[0cm]{0cm}{0cm}
$\sigma_{tot}$ (fb)}
\\ \hline
\rule[0cm]{0cm}{0cm}
$\mu$ (GeV) & $\sqrt s_{pp}=14$ TeV \\ \hline\hline
\rule[0cm]{0cm}{0cm}
$2m_t$ & $10474\pm38(1856.3\pm8.6)$ \\ 
400    & $10030\pm36(1719.2\pm6.9)$ \\ 
500    & $9459\pm32(1505.8\pm6.2)$ \\ 
600    & $8913\pm30(1356.4\pm5.6)$ \\ 
700    & $8539\pm27(1242.3\pm5.1)$ \\ 
800    & $8258\pm24(1153.4\pm4.9)$ \\ 
900    & $8019\pm22(1082.5\pm4.6)$ \\ 
1000   & $7802\pm19(1010.3\pm4.4)$ \\ \hline\hline
\multicolumn{2}{|c|}
{\rule[0cm]{0cm}{0cm}
$M_H=150$ GeV}
\\ \hline

\multicolumn{2}{|c|}
{\rule[0cm]{0cm}{0cm}
no acceptance cuts}
\\ \hline

\multicolumn{2}{|c|}
{\rule[0cm]{0cm}{0cm}
MRRS(3)}
\\ \hline

\multicolumn{2}{c}
{\rule{0cm}{1.0cm}
{\Large Tab. II}}  \\

\end{tabular}
\end{center}
\end{table}

\vfill
\newpage 

\begin{table}
\begin{center}
\begin{tabular}{|c|c|c|}
\hline
\multicolumn{3}{|c|}
{\rule[0cm]{0cm}{0cm}
$\sigma_{tot}$ (fb)}
\\ \hline
\rule[0cm]{0cm}{0cm}
$m_{d_4}/m_t$ (GeV) & $bg\ar bd_4\bar{d}_4$ & $bg\ar bt\bar t$\\ \hline\hline
\rule[0cm]{0cm}{0cm}
165 & $7112\pm27$ & $11303\pm34$ \\ 
170 & $6142\pm25$ & $10062\pm27$ \\ 
175 & $5319\pm25$ & $8101\pm24$ \\ 
180 & $4626\pm23$ & $8093\pm20$ \\ 
185 & $4029\pm17$ & $7305\pm18$ \\ \hline\hline
\multicolumn{3}{|c|}
{\rule[0cm]{0cm}{0cm}
$M_H=150$ GeV}
\\ \hline

\multicolumn{3}{|c|}
{\rule[0cm]{0cm}{0cm}
no acceptance cuts}
\\ \hline

\multicolumn{3}{|c|}
{\rule[0cm]{0cm}{0cm}
MRRS(3)}
\\ \hline

\multicolumn{3}{|c|}
{\rule[0cm]{0cm}{0cm}
$\sqrt s_{pp}=14$ TeV}
\\ \hline

\multicolumn{3}{c}
{\rule{0cm}{1.0cm}
{\Large Tab. III}}  \\

\end{tabular}
\end{center}
\end{table}

\vfill
\newpage 

\begin{figure}[p]
~\epsfig{file=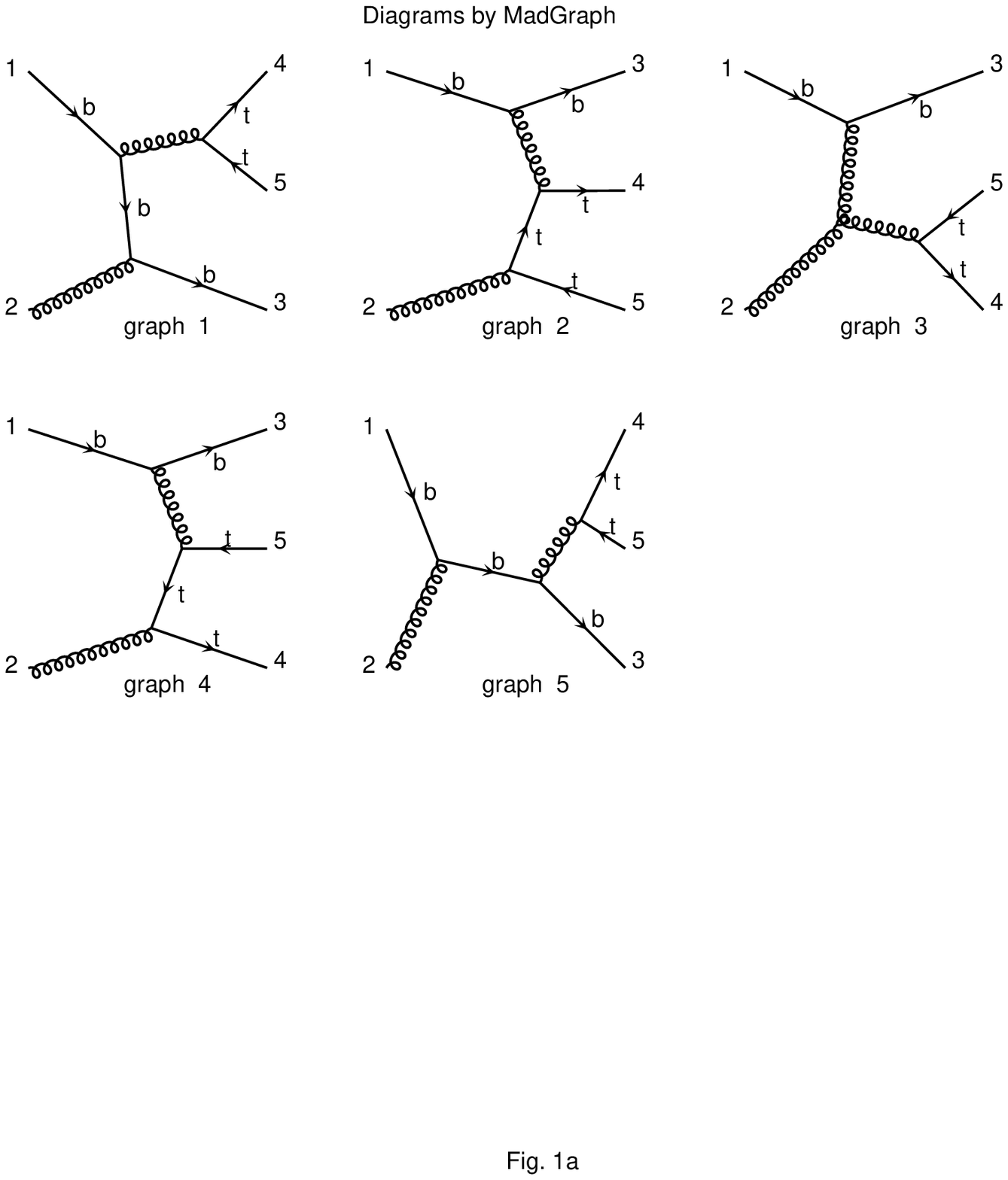,height=22cm}
\vspace*{2cm}
\end{figure}
\vfill
\clearpage

\begin{figure}[p]
~\epsfig{file=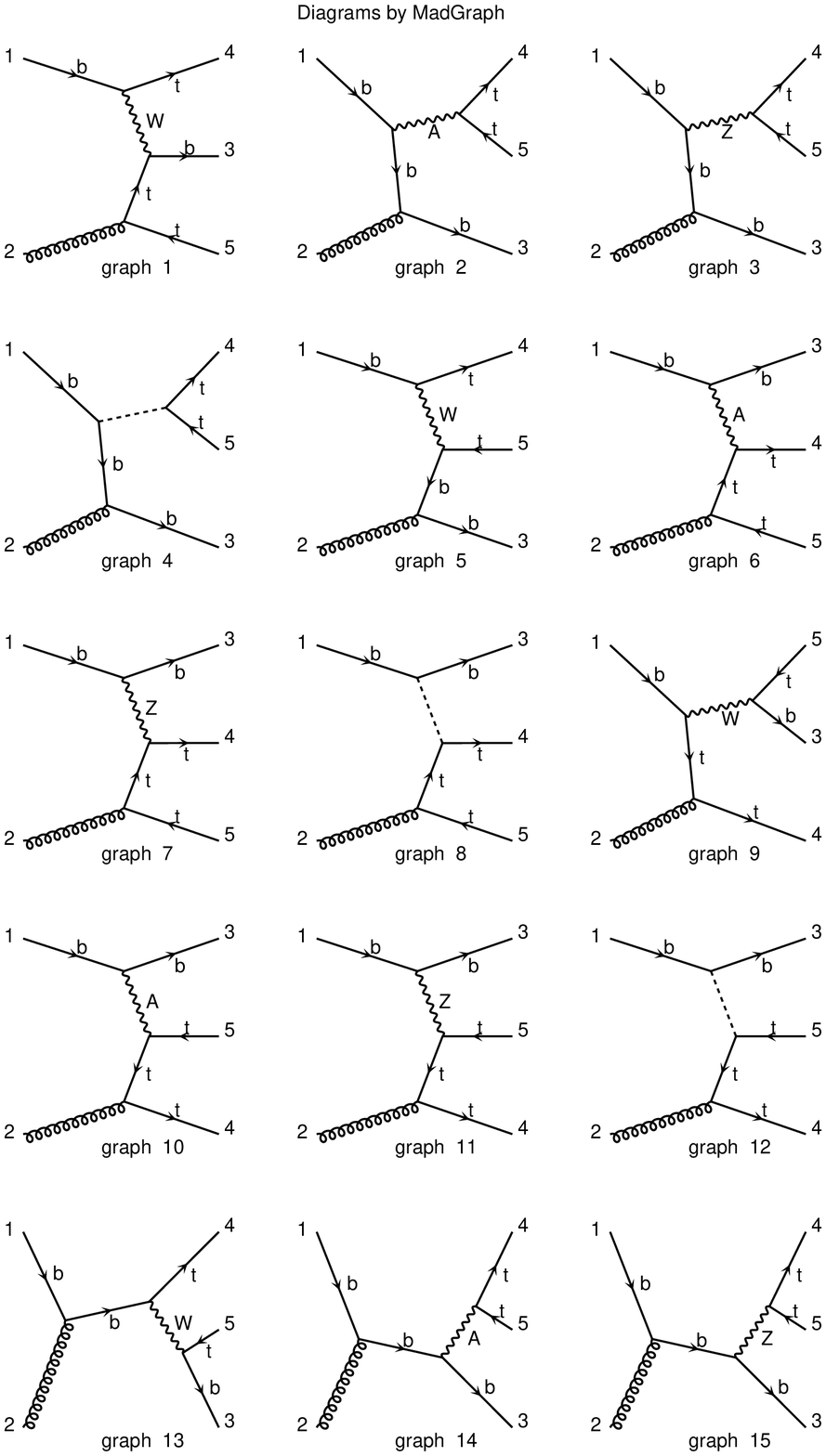,height=22cm}
\vspace*{2cm}
\end{figure}
\vfill
\clearpage

\begin{figure}[p]
~\epsfig{file=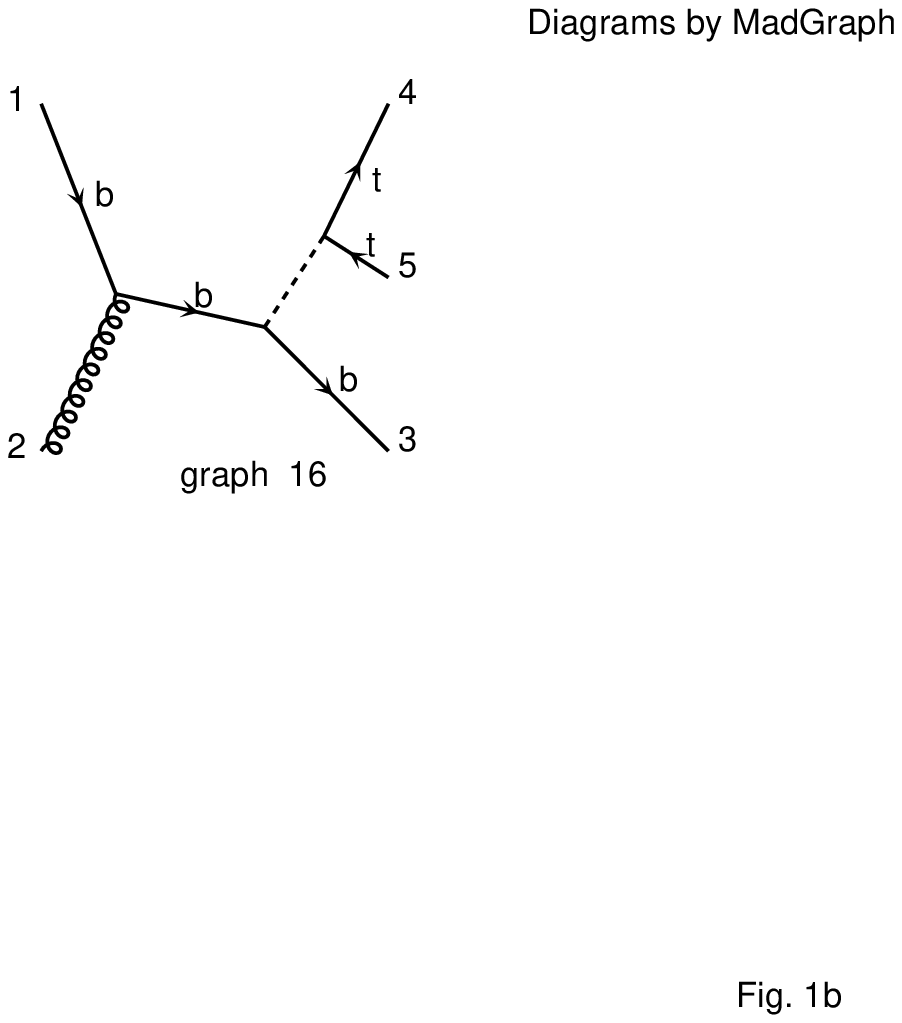,height=22cm}
\vspace*{2cm}
\end{figure}
\vfill
\clearpage

\begin{figure}[p]
~\epsfig{file=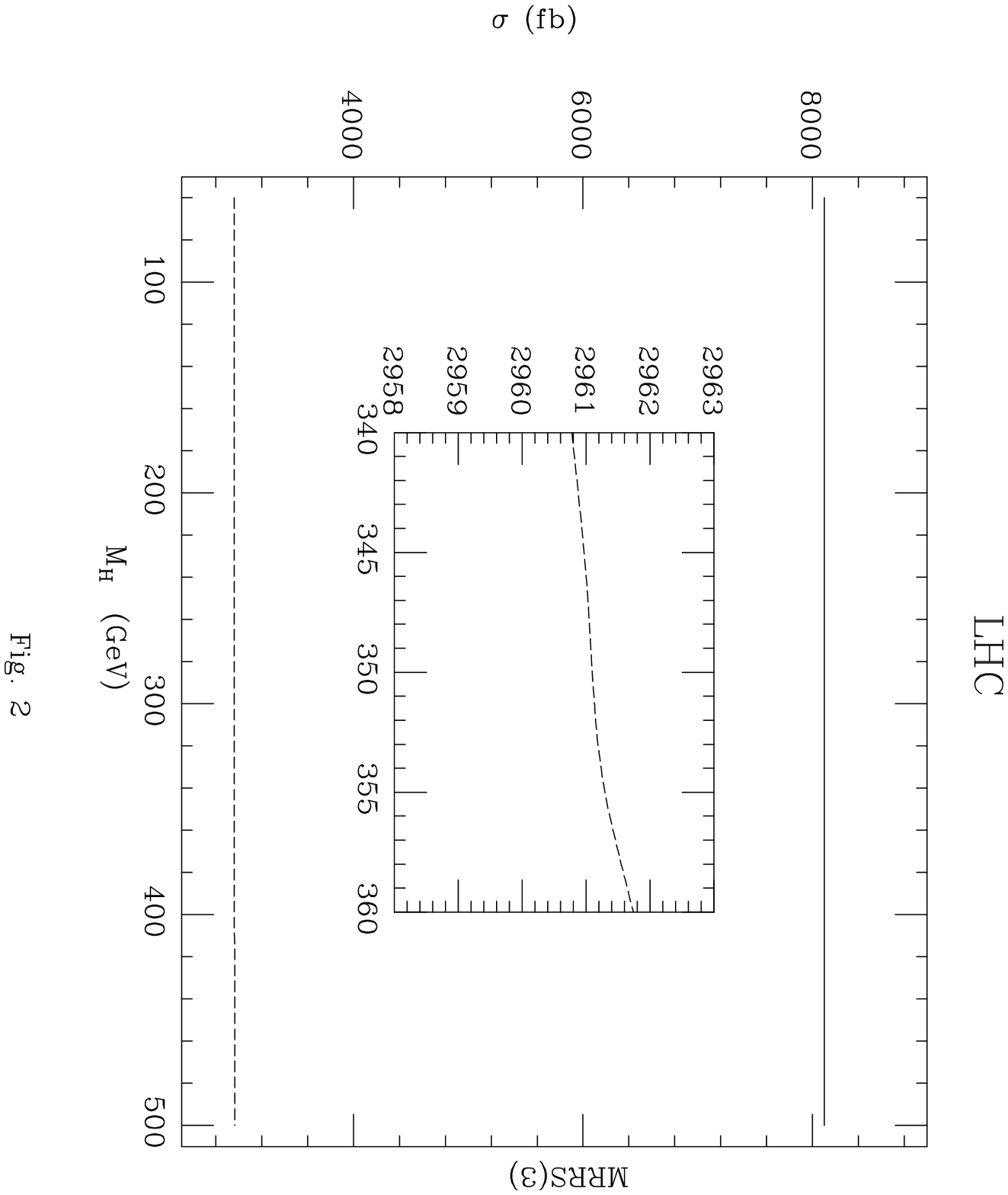,height=22cm}
\vspace*{2cm}
\end{figure}
\vfill
\clearpage

\begin{figure}[p]
~\epsfig{file=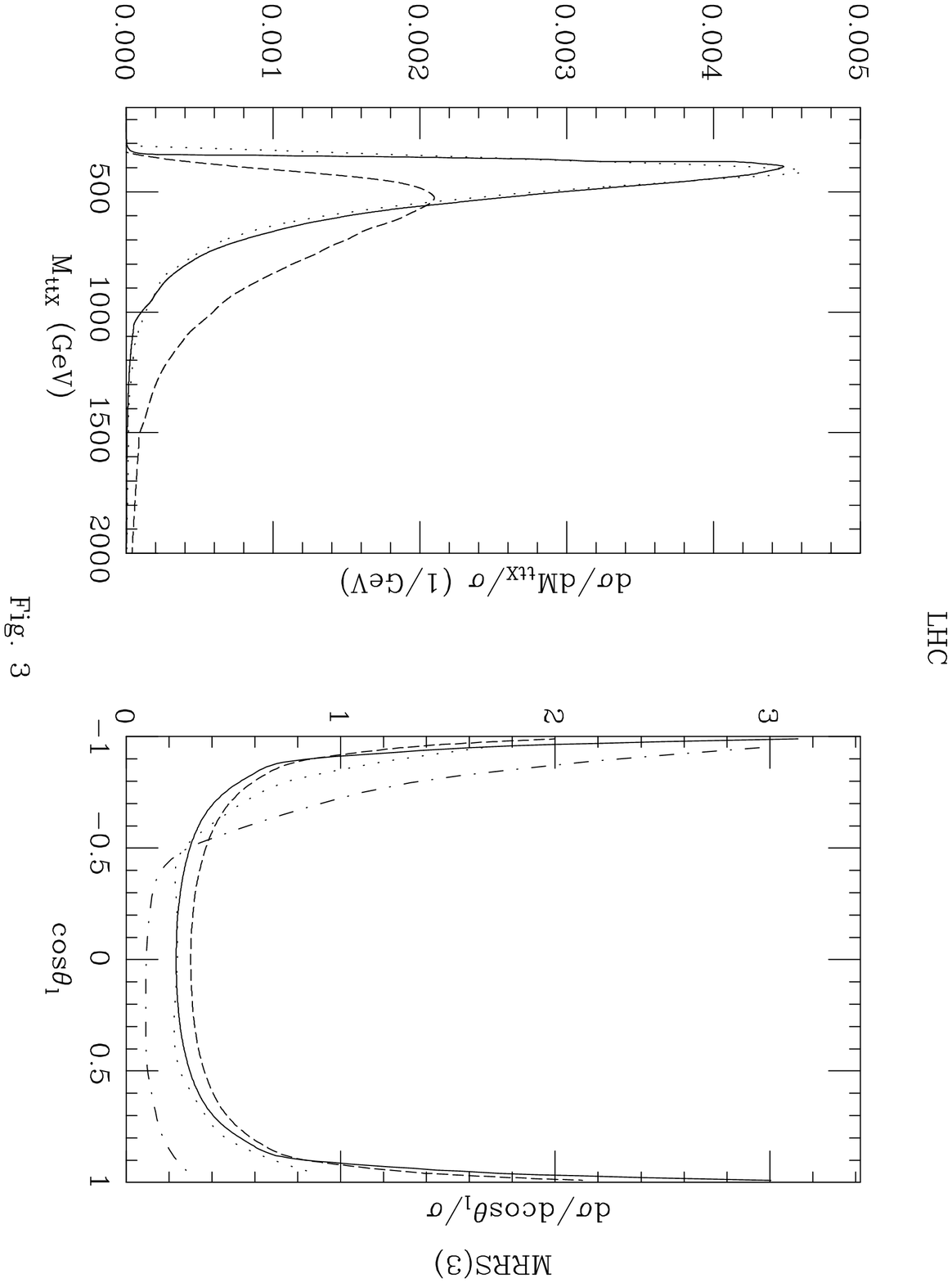,height=22cm}
\vspace*{2cm}
\end{figure}
\vfill
\clearpage

\begin{figure}[p]
~\epsfig{file=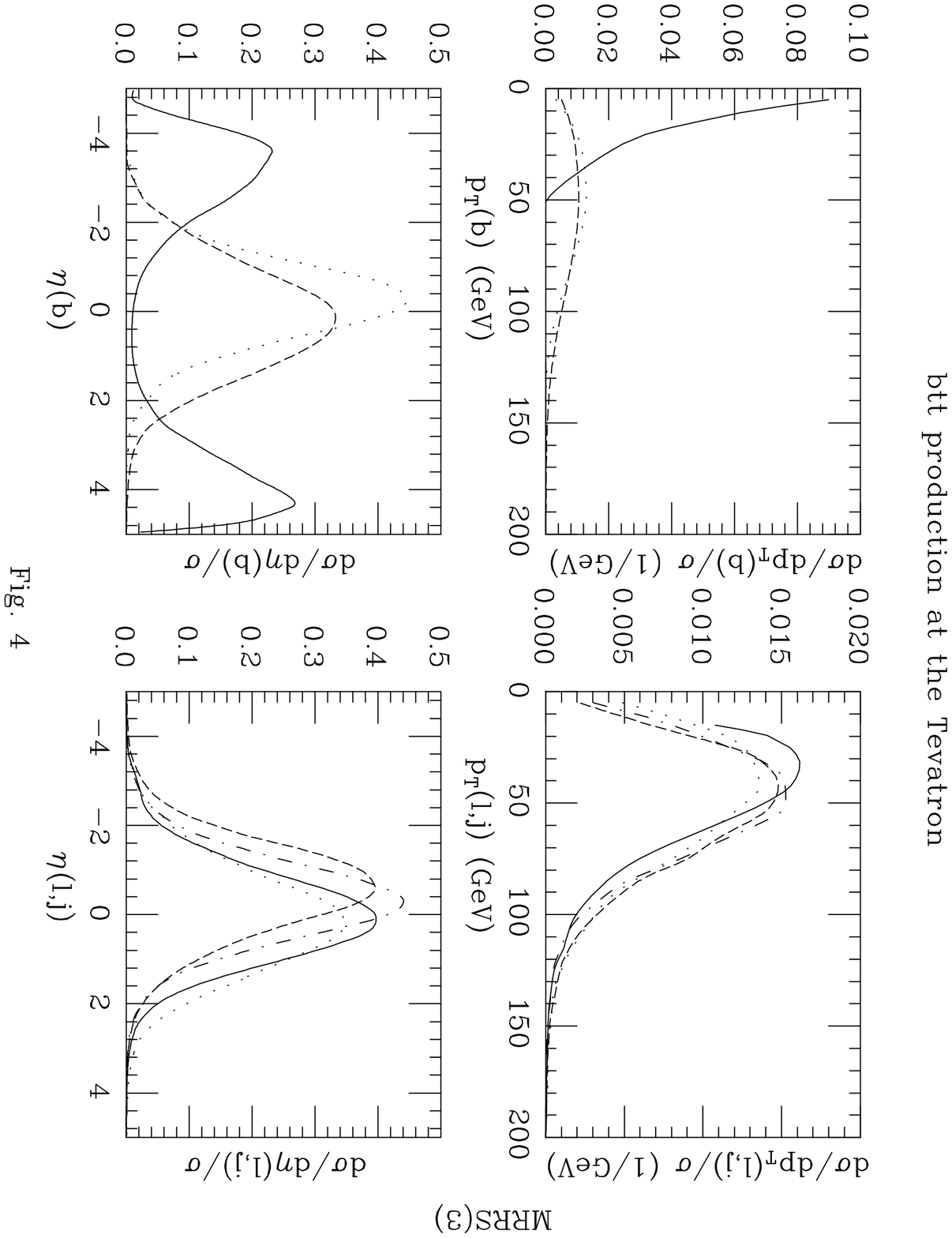,height=22cm}
\vspace*{2cm}
\end{figure}
\vfill
\clearpage

\begin{figure}[p]
~\epsfig{file=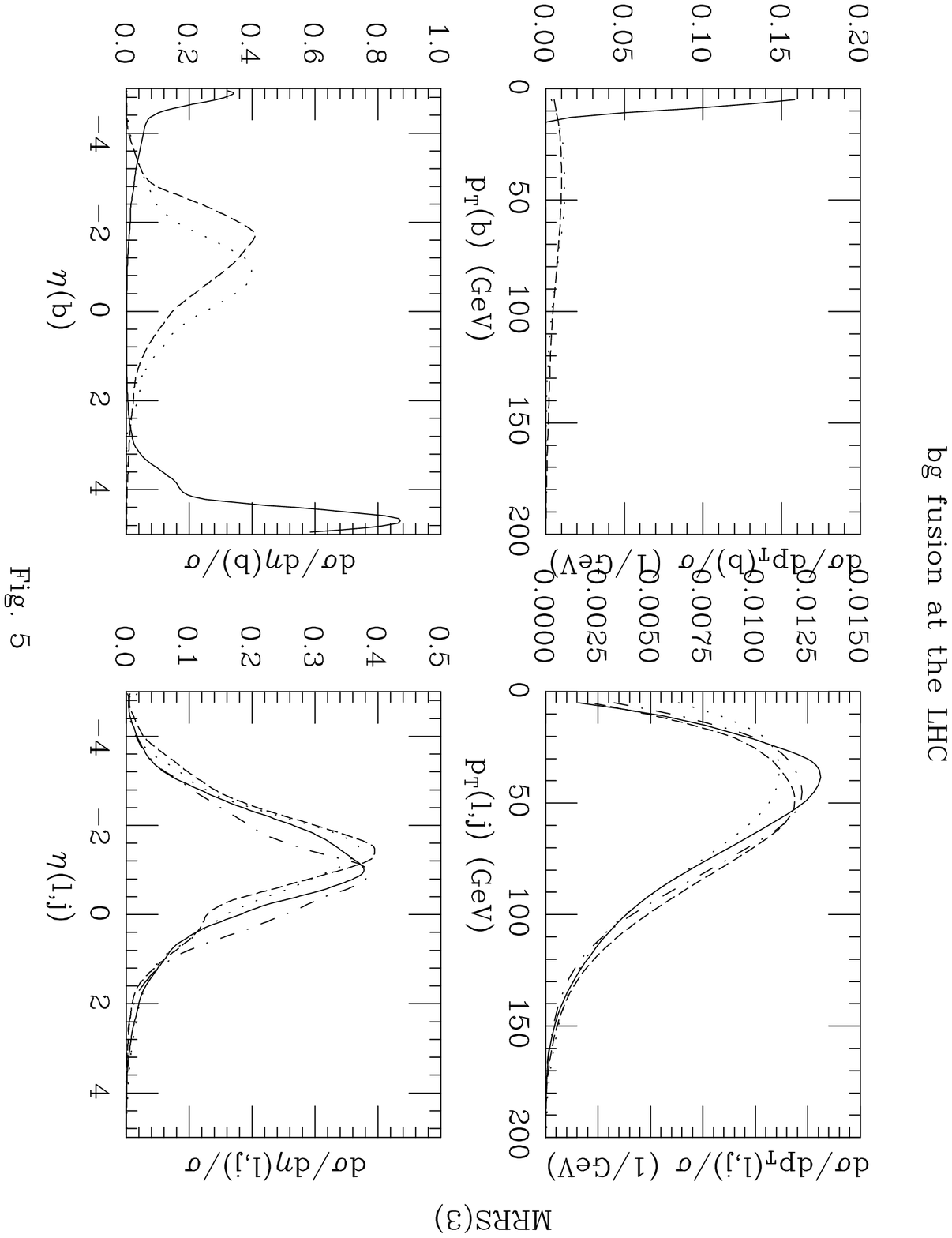,height=22cm}
\vspace*{2cm}
\end{figure}
\vfill
\clearpage

\vfill
\end{document}